\documentclass[ preprintnumbers,A4paper,preprint,prb]{revtex4}
\usepackage{graphicx}

\newcommand{\ket}[1]{| #1 \rangle}

\begin{document}

\centerline {\Large \bf Quantum Computing in Arrays}

\centerline {\Large \bf Coupled by `Always On' Interactions}

\bigskip

\centerline {{\bf S. C. Benjamin$^{1,2}$ and S. Bose$^{3}$}}

\smallskip

{\footnotesize

\centerline {$^1$Ctr. for Quantum Computation, Clarendon Lab., Univ. of Oxford, {\scriptsize OX1 3PU}, UK.}

\centerline {$^2$Dept. of Materials, Parks Road, Univ. of Oxford,  {\scriptsize OX1 3PH},  UK.}

\centerline {$^3$Dept. of Physics and Astronomy, University College London,}

\centerline{Gower St., London {\scriptsize WC1E 6BT}, UK.}

}

\smallskip

{\bf It has recently been shown that one can perform quantum computation in a Heisenberg chain in which the interactions are `always on', provided that one can abruptly tune the Zeeman energies of the individual (pseudo-)spins. Here we provide a more complete analysis of this scheme, including several generalizations. We generalize the interaction to an anisotropic  form (incorporating the XY, or Forster, interaction as a limit), providing a proof that a chain coupled in this fashion tends to an effective Ising chain in the limit of far off-resonant spins. We derive the primitive two-qubit gate that results from exploiting abrupt Zeeman tuning with such an interaction. We also demonstrate, via numerical simulation, that the same basic scheme functions in the case of smoothly shifted Zeeman energies. We conclude with some remarks regarding generalisations to two- and three-dimensional arrays.}

\smallskip

There has recently been considerable interest in the question of whether one can perform quantum computation (QC) in Heisenberg-type systems (e.g. interacting electron spins) when the interaction is `always-on' \cite{ourPRL, zhou, newLANL}. This question follows on from a work concerning Heisenberg systems in which the interactions are presumed to be switchable, either individually\cite{3qubitExchangeOnly, Levy, myABqubitPaper} or collectively\cite{ababPRL}. Numerous proposals exist\cite{DiVincenzo1,kane,spinResTrans} for experimental realization of such a model, however interaction switching is liable to prove very challenging to realize, and this motivates the interest in `always-on' interactions. In Ref. [\onlinecite{ourPRL}] we proposed a scheme for exploiting a simple one-dimensional Heisenberg chain with constant, isotropic nearest neighbor interactions. The scheme involved adjusting the single-spin level splittings (the Zeeman energies) to bring neighbors in and out of resonance with one another. We exploited the fact that far off-resonance spins do not exchange energy, but rather interact in an Ising `ZZ' form. We argued that by separating the qubit-bearing spins by passive `barrier' spins, one can negate this residual interaction (thus achieving a passive state for the array) - yet one can invoke an interaction on demand simply by bringing a barrier into resonance with its neighbors. 

In the present paper we elaborate on several aspects of that earlier Letter, and we provide certain extensions. Whereas previously we considered only one specific form for the interaction, i.e. the isotropic Heisenberg form ($\sigma^X\sigma^X+\sigma^Y\sigma^Y+\sigma^Z\sigma^Z$), we now generalize our arguments to accommodate different magnitudes for the in-plane and perpendicular components. Thus we subsume the prior isotropic form, and the purely planar ``XY'' interaction, as special cases. There is a wide variety of promising physical systems associated with this family of interactions (for the isotropic limit, see e.g. Refs. [\onlinecite{DiVincenzo1,kane,spinResTrans}], and for the anisotropic case, Refs. [\onlinecite{Imamoglu,Mozyrsky,Seiwert}]). The XY limit is also referred to as the Forster interaction, especially when studied in the context of excitonic exchange in biological molecules. With this generalized form of interaction, we first present an analysis of the effect of far off-resonant neighbors in a long chain, obtaining the anticipated Ising-like form as the lowest order term. We then explain in detail how Zeeman tuning can be exploited perform an elementary two-qubit gate,  and we show how the resulting unitary operation depends on the Z versus XY asymmetry in the interaction.

Whereas the original paper assumed a perfectly abrupt transition between on-resonant and far off-resonant Zeeman energies, here we follow our analysis with a numerical simulation demonstrating that smoothly changing Zeeman energies can implement the gate process equally well. This observation considerably increases the practicality of the scheme. Finally, we discuss the generalization to two- and three-dimensional arrays.

\bigskip\bigskip
\noindent{\bf Analysis of Heisenberg Chain with Large Zeeman Discrepancies}

The analysis is presented in full in Appendix I. Here we summarize it. We start from a total Hamiltonian $H$ given by
$$
H=H_{single}+H_{int}, 
$$
where
$$
H_{single} = \sum_j B_j \sigma_i^Z. \nonumber
$$
and the exchange interaction is as follows, where the factor $\alpha$ allows for a possible anisotropy between the in-plane and z-direction components.
\begin{eqnarray}
H_{int}&=&J\sum_j (\sigma^X_j\sigma^X_{j+1}+\sigma^Y_i\sigma^Y_{j+1}+\alpha\sigma^Z_j\sigma^Z_{j+1})\nonumber \\
&=&\frac{J}{2}(\sum_j \sigma^{+}_j
\sigma^{-}_{j+1}+\sigma^{-}_j\sigma^{+}_{j+1})+J\alpha\sum_j\sigma_j^Z\sigma_{j+1}^Z, \nonumber
\end{eqnarray}
where $\sigma^\pm\equiv\sigma^X\pm i\sigma^Y$. Here and below, the sum ranges over all $N$ qubits, but subscripts such as $j+1$ are understood to be modulo $N$, i.e. we {\bf assume a closed circular topology}. This considerably simplifies the analysis, but it is not a real constraint - in the limit of large chains the open and closed topologies will be equivalent.

We rewrite $H=H_{1}+H_{2}$ where

$$
H_{1} = \sum_j B_j \sigma_j^Z+J\alpha\sum_j\sigma_j^Z\sigma_{j+1}^Z. \nonumber
$$
and
$$
H_{2}=\frac{J}{2}(\sum_j \sigma^{+}_j
\sigma^{-}_{j+1}+\sigma^{-}_j\sigma^{+}_{j+1}). \nonumber
$$
Notice that $H_1$ is simply the Hamiltonian for an Ising spin chain with varying Zeeman energies. We will find that this term dominates the time evolution when the spins are far off-resonance with their neighbors; the contribution of $H_2$ then vanishes.

Our approach is to exploit the Trotter formula to manipulate the time evolution operator into a form that can be recognized as Ising and non-Ising parts. This is detailed in Appendix I. The exact expression for the time evolution is found to be:
\begin{eqnarray}
U(t)=R(t)\exp{(-iH_1 t)}\ \ \ \ {\rm where}& R=\{\prod_{m=1}^{m=n}\exp\lgroup\frac{-it}{n}H_R(\frac{mt}{n})\rgroup\}_{n\rightarrow\infty}\nonumber\\
H_R(\eta)=\sum_j X_j(\eta)\sigma^+_j\sigma^-_{j+1}+X^\dagger_j(\eta)\sigma^-_j\sigma^+_{j+1}\ \ &{\rm with}\ \ X_j(\eta)=\exp(i\eta(\Delta_j+2J\alpha(\sigma^Z_{j+2}-\sigma^Z_{j-1}))\nonumber
\end{eqnarray}

\noindent where $\Delta_j\equiv 2(B_{j+1}-B_j)$. The right hand term in $U(t)$ is the pure Ising chain evolution we seek, but the subsequent `residual' operator $R$ is more complex. In the second part of Appendix II we expand $R$ as a power series and inspect the terms. We conclude that, for a regular chain with a characteristic $\Delta$ (such as an $ABABAB..$ chain where $\Delta_j=(-1)^j\Delta$), the time evolution can be written as 
$$
U(t)=(1-\delta P(t))\exp{(-iH_1 t)} 
$$
where $\delta\equiv J/\Delta$, for some finite operator $P(t)$ whose magnitude does not increase with $\Delta$. {\bf Thus for any given time period $t$ the non-Ising evolution will be negligible if $\Delta$ is {\em sufficiently} large compared to $J$.} Assuming that we can dynamically change a $\Delta_j$, switching it between zero and a large value, we can then exploit this result to produce a form of `gate' for quantum computation.

\bigskip\bigskip
\noindent {\bf Exploitation of the Heisenberg-to-Ising Transition to Perform QC}
\bigskip

Assume that we have some array in which every pair of adjacent spins is far off resonance from one another, i.e. $\Delta_j\gg J$, $\forall j$. Now assume that we abruptly tune one (or more) of the spin Zeeman energies so that we have a triplet $ABA$ where energies $A$ and $B$ are comparable. Let us refer to these spins by the labels $1$ to $3$, and similarly label the external neighboring spins as $0$ and $4$. Suppose spins $0$, $2$ and $4$ are initially in state $\ket{\uparrow}$. Since spin $0$ remains far off resonance from $1$, their interaction is effectively of the Ising form $J\alpha\sigma_0^Z\sigma_1^Z$. Similarly the interaction between $3$ and $4$ is $J\alpha\sigma_3^Z\sigma_4^Z$. Moreover, those external spins (having only an Ising interaction with their neighbors) are `frozen' in the $\ket{\uparrow}$ state thus their interaction with the triplet reduces to $J\alpha\sigma_1^Z$ and $J\alpha\sigma_3^Z$, and the dynamics of the triplet are described by the Hamiltonian:
$$
H_{\rm triplet}=H_{\rm zeeman}+H_{\rm int}
$$
$$
H_{\rm zeeman}=(A+\alpha J)(\sigma_1^Z+\sigma_3^Z)+B\sigma_2^Z
$$
$$
H_{\rm int}=J\sum_{j=1,2}\sigma_j^X\sigma_{j+1}^X+\sigma_j^Y\sigma_{j+1}^Y+\alpha\sigma_j^Z\sigma_{j+1}^Z
$$
In the following we will use the notation $J_{XY}\equiv J$, $J_Z\equiv \alpha J$, $a \equiv A+J_Z$ (the effective Zeeman energy of spins 1 and 3) and $b \equiv B$ for consistency. The Hamiltonian is easy to analyze; the states $\ket{\uparrow\uparrow\uparrow}$ and $\ket{\downarrow\downarrow\downarrow}$ of course remain eigenstates while the remaining states form two distinct subspaces. For the `up' subspace spanned by $\{\ket{\downarrow\uparrow\uparrow}$, $\ket{\uparrow\downarrow\uparrow}$, $\ket{\uparrow\uparrow\downarrow}\}$ we have Hamiltonian and eigenvectors given by
$${\hat H}_U= b\ {\bf I} + 2J_{XY}\left( 
\begin{array}{ccc}
0 & 1 & 0 \\ 
1 & p &  1 \\ 
0 & 1 &  0 
\end{array}
\right)\ \ \Rightarrow\ \ \ 
\ket{a}_U=\left( 
\begin{array}{c}
1  \\ 
0  \\ 
-1 
\end{array}
\right)\ \ \ {\rm and}\ \ \ 
\ket{\pm}_U=\left( 
\begin{array}{c}
1 \\ 
\frac{1}{2}(p\pm S_p) \\ 
1
\end{array}
\right).
$$
With corresponding energies $E_a^U=b$, $E_\pm^U=b+J_{XY}(p\pm S_p)$. Here $p\equiv (a-b-J_Z)/J_{XY}$ and $S_p\equiv\sqrt{8+p^2}$.
Similarly for the complimentary `down' space $\{\ket{\uparrow\downarrow\downarrow}$, $\ket{\downarrow\uparrow\downarrow}$, $\ket{\downarrow\downarrow\uparrow}\}$ we have
$${\hat H}_D= -b\ {\bf I} + 2J_{XY}\left( 
\begin{array}{ccc}
0 & 1 & 0 \\ 
1 & q &  1 \\ 
0 & 1 &  0 
\end{array}
\right)\ \ \Rightarrow\ \ \ 
\ket{a}_D=\left( 
\begin{array}{c}
1  \\ 
0  \\ 
-1 
\end{array}
\right)\ \ \ {\rm and}\ \ 
\ket{\pm}_D=\left( 
\begin{array}{c}
1 \\ 
\frac{1}{2}(q\pm S_q) \\ 
1
\end{array}
\right).
$$
With energies $E_a^D=-b$, $E_\pm^D=b+J_{XY}(p\pm S_p)$, where $q\equiv (b-a-J_Z)/J_{XY}$ and $S_q\equiv\sqrt{8+q^2}$. Now, we know that the initial computational qubit states are 
$$
\begin{array}{ccc}
\ket{00}  =& \ket{\downarrow\uparrow\downarrow}& \ \ \Leftarrow {\rm\ composed\ of }\ \ket{+}_D\ {\rm and}\ \ket{-}_D\ \ \ \ \ \ \\
\ket{01} = & \ket{\downarrow\uparrow\uparrow}& \ \ \Leftarrow {\rm\ composed\ of }\ \ket{a}_U, \ket{+}_U\ {\rm and}\ \ket{-}_U\\
\ket{10} = & \ket{\uparrow\uparrow\downarrow}& \ \ \Leftarrow {\rm\ composed\ of }\ \ket{a}_U, \ket{+}_U\ {\rm and}\ \ket{-}_U\\
\ket{11} = & \ket{\uparrow\uparrow\uparrow}& \ \ \Leftarrow {\rm \ eigenstate\ \ \ \ \ \ \ \ \ \ \ \ \ \ \ \ \ \ \ \ \ \ \ \ \ \ \ \ \ \ }
\end{array}
$$
During the gate operation, the states (other than $\ket{11}$) will rotate within their subspaces. We must arrange to `revive' both the $\ket{00}$ state and the states $\ket{01}$ \& $\ket{10}$ at the same instant, i.e. we must arrange that at some time $t_R$ the central spin is in the definite state $\ket{\uparrow}$ for all computational basis states. (Note that this condition does permit a net rotation in the plane defined by $\ket{01}$ \& $\ket{10}$). Thus at that moment we can effectively switch off the exchange interaction (by switching to far off-resonant Zeeman energies) and we will have performed some unitary transform in the computational basis. Whether such a transform constitutes a useful gate depends on entanglement criteria as mentioned later.
The times for which $\ket{00}$ revives are determined by $E_+^D-E_-^D$. The times at which a state, initially in the $\ket{01}$, $\ket{10}$ plane, returns to that plane are determined by $E_+^U-E_-^U$. Now the parameter which we can experimentally vary is the Zeeman detuning $a-b$; although there may be various detunings for which the revivals coincide (which could be found numerically), there is one value that is immediately obvious by inspection: $a-b=0$ (corresponding to tuning the central barrier spin to $A+J_Z$). In this case we see that $p=q=-J_Z/J_{XY}$, $S_p=S_q=\sqrt{8+(J_Z/J_{XY})^2}$ and thus both revivals coincide at time $t_R=\pi\hbar(8J_{XY}^2+J_Z^2)^{-\frac{1}{2}}$. At this instant, the transformation in the computational basis $\{\ket{00}$, $\ket{01}$, $\ket{10}$, $\ket{11}\}$ is given by the following matrix (neglecting a global phase)
$$
U=\left(
\begin{array}{cccc}
1 &0 &0 &0\\
0 &iQs &Qc &0\\
0 &Qc &iQs &0\\
0 &0 &0 &W
\end{array}
\right)
$$
Here $Q=-\exp(i\phi)$, $s/c=\sin/\cos(\phi)$ and $W=-\exp(-2i\phi)$ with $\phi=\frac{\pi}{2}(8J_{XY}^2/J_Z^2+1)^{-\frac{1}{2}}$. The phases in this matrix are with respect to the passive state of the device (i.e. if we had not tuned the triplet into resonance), under the assumption that the resonance was achieved by shifting the Zeeman energy of the central spin. (If in fact the Zeeman energies of the qubit-bearing spins were adjusted to achieve resonance, then we simply have the above matrix together with two trivial single qubit $Z$ gates.) This transformation $U$ is entangling, and is therefore adequate to construct a universal gate set when combined with single qubit gates\cite{Nielsen}. Using the procedure described in Refs. [\onlinecite{Nielsen, gatePaper}] one can confirm that no more than four uses of this gate are required to form a Control-NOT, for a wide range of $J_Z$ including the $J_Z=0$ and $J=J_Z$ cases, which represent the XY interaction and the isotropic Heisenberg interaction, respectively. It is easier to appreciate the nature of the transform if we apply a couple of single-qubit Z-rotations; defining
$$
Z(\theta)\equiv\left(
\begin{array}{cc}
\exp (i \theta) &0\\
0 & \exp ( -i \theta ) 
\end{array}\right)
$$
then neglecting a global phase,
\begin{equation}
Z_1(\psi).Z_2(\psi).U=
\left(
\begin{array}{cccc}
1 &0 &0 &0 \\
0 &-Q's &iQ'c &0\\
0 &iQ'c &-Q's &0\\
0 &0 &0 &1
\end{array}
\right)
\label{dressedMat}
\end{equation}
Here $\psi=\frac{\pi}{4}(1-(8J^2/J_Z^2+1)^{-\frac{1}{2}})$ and $Q'=Q^2$ while $s/c$ are as before.
Notice that for the $J_Z=0$ limit, i.e. the case of a pure XY interaction, then the primitive matrix $U$ takes a particularly simple form \cite{recentConfirmation}
\begin{equation}
U_P=\left(
\begin{array}{cccc}
1 &0 &0 &0\\
0 &0 &-1 &0\\
0 &-1 &0 &0\\
0 &0 &0 &-1
\end{array}
\right)
\label{XYgate}
\end{equation}
using which one can construct a CNOT with only two applications, as shown in Fig. 1(b). In this limit, the dressed matrix (\ref{dressedMat}) is recognizable as the ``iSWAP'' which has been studied in the context of an XY interaction between adjacent qubits\cite{Schuch}. Indeed, in the limit of a {\em strict} XY interaction, one might choose to abandon the barrier spin architecture completely, and adopt a trivial architecture in which qubits are adjacent (since the primary function of the barrier spins is to negate the effect of the residual Ising interaction, absent for the pure XY form). 

\begin{figure}[!t]
\centering
\resizebox{10.7cm}{!}{\includegraphics{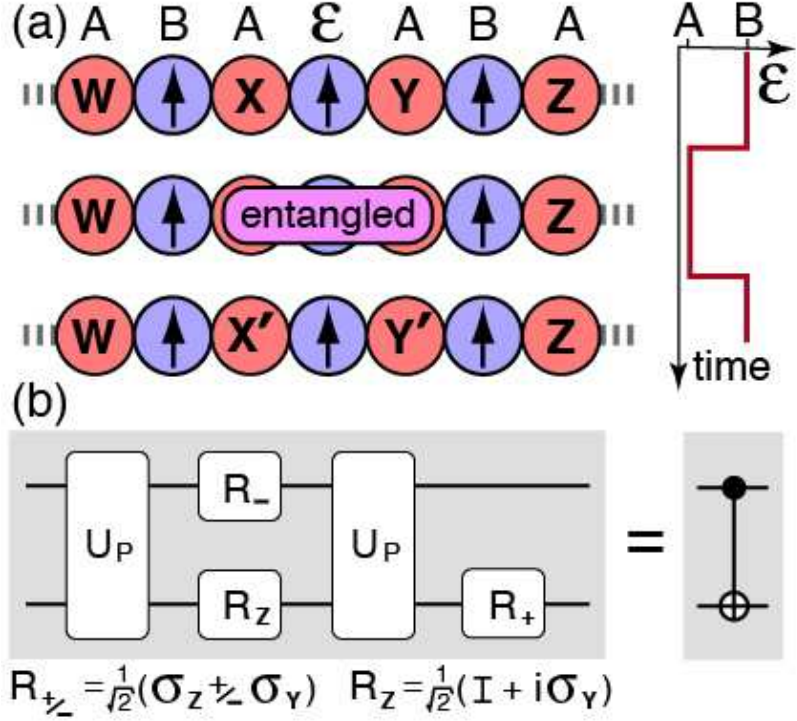}}

\vspace{0.2cm}

\caption{(a) Schematic showing the basic two-qubit gate. Letters $A$, $B$, {\Large $\epsilon$} denote Zeeman energies; $A$ and $B$ are fixed but {\Large $\epsilon$} is abruptly changed as shown by the graph to the right. The operation is completed when the central spin `revives' into the state $\ket{\uparrow}$. The resulting primitive two-qubit gate is entangling for all values of the anisotropy parameter $\alpha\equiv J_Z/J_{XY}$. In the special case of $J_Z=0$ the gate has a particularly simple form (eqn. \ref{XYgate}) and a corresponding circuit for the control-NOT operation (b) requires only two such primitives. }

\label{figure1}

\end{figure}

\begin{figure}[!t]
\centering
\resizebox{10.7cm}{!}{\includegraphics{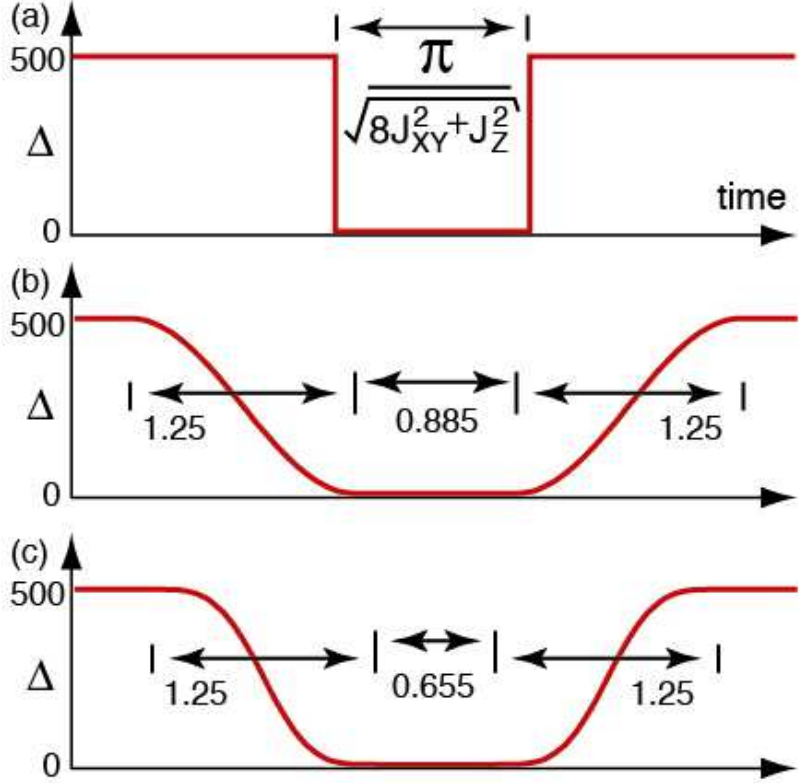}}

\vspace{0.2cm}

\caption{(a) An abrupt Zeeman shift, corresponding to our analytic treatment. (b) \& (c) Numerical simulation demonstrating that other, smooth functions can also suffice. In (b) we use a `gentle' switching function of the form $\cos^2\left(\pi\frac{t-t_0}{2 t_\Delta}\right)$, where $t_0$ is the time at which the switching process begins and $t_\Delta =1.25$ is the switch duration. After a `flat' period, the function is reversed to return us to the large `passive' detuning. In (c) we construct a $4^{\rm th}$ power sinusoidal function for a sharper profile \cite{sin4function}. In each case we find that simply choosing the correct duration for the central constant phase allows complete\cite{completeR} barrier revival. In these graphs, the time axis is in units of $\hbar/J_{XY}$, and we arbitrarily chose an anisotropic interaction with $J_Z/J_{XY}=0.7$.}

\label{smoothFigure}

\end{figure}

\begin{figure}[!t]
\centering
\resizebox{10.7cm}{!}{\includegraphics{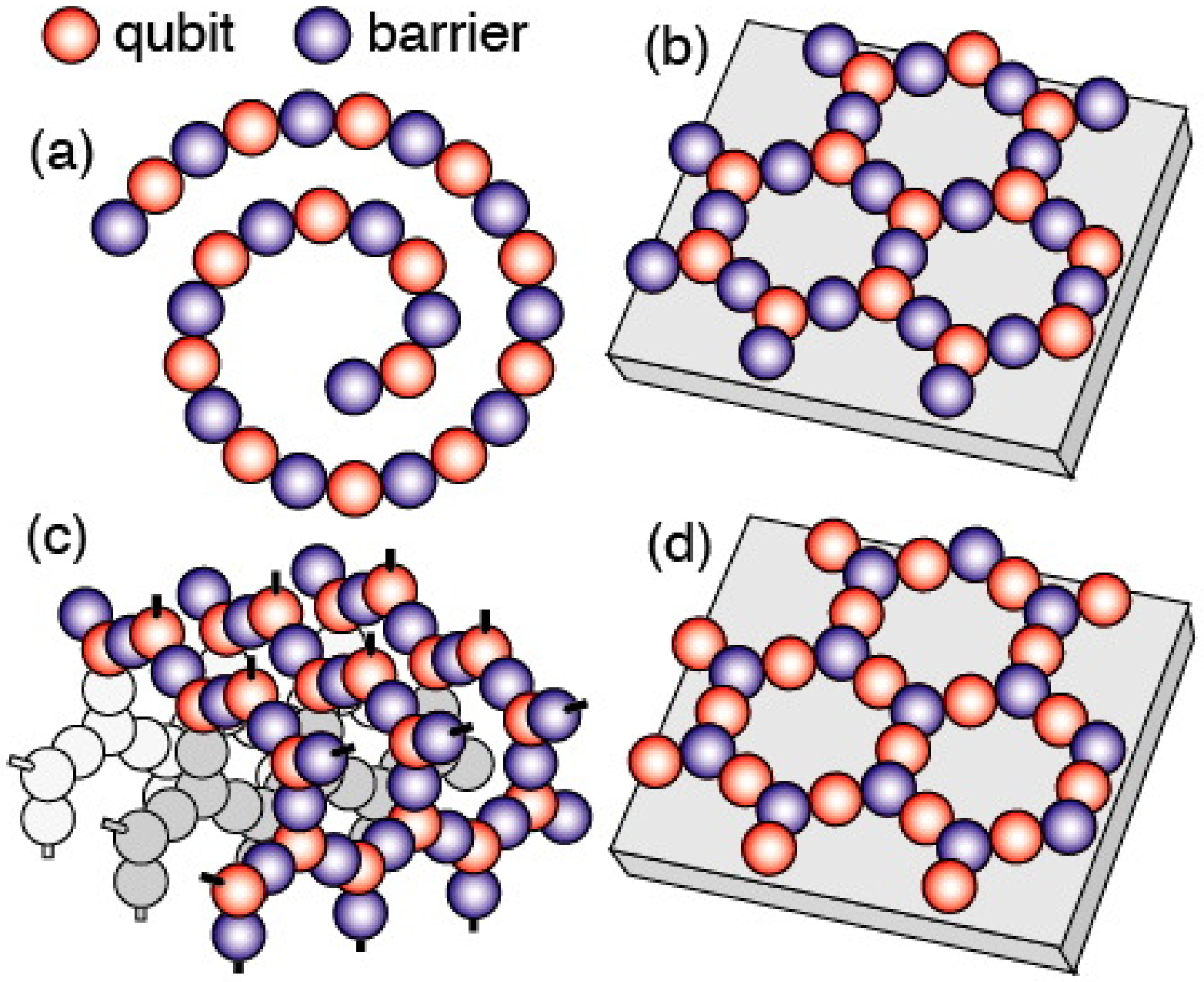}}

\caption{Structures that are efficient in terms of $R_Q$, the ratio of number of qubits stored to {\em total} number of spins. In one dimension, using the gate process analysed here (c.f. Fig. 1) the value of $R_Q$ is necessarily $1/2$. For higher dimensions, if one employs the same gate process (by `inserting' a unique barrier between each pair of adjacent qubits) then the best value of $R_Q$ for a regular structure is $2/5$; structures meeting this limit are shown in (b) and (c). If, however, one allows each barrier to separate multiple qubits, then higher ratios are possible - for (d) $R_Q=3/5$.}

\label{figure3}

\end{figure}

Note the second form of gate presented in Ref.[\onlinecite{ourPRL}] can also be generalised to anisotropic Heisenberg interactions, although it does require some finite $Z$ component since this is exploited to accumulate a phase during the gate operation. 

In Ref.[\onlinecite{ourPRL}] and in the above analysis, we consider an abrupt change from far off-resonance spins into resonance. This may be difficult to achieve in many otherwise promising implementations, therefore it we now investigate the effect of smooth switching. Figure 2 shows various profiles for the dynamically changing Zeeman energy of the central spin, given that the Zeeman energies of the outer spins are static (Fig. 2(a) corresponds to analytic treatment given above). For numerical convenience we have built these switching profiles as piece-wise combinations of analytic functions, as defined in the figure caption. In both cases (b) and (c) we fixed the time for the switching transition to the arbitrary choice $t_\Delta=1.25$ and varied just a single parameter, the time for which the detuning is zero. The values shown in the Figure provided a complete revival of the central spin for all qubit basis states, just as in the case of the abrupt transition. The specific transformation achieved in the qubit basis (i.e. the analogue of eqn. (\ref{dressedMat})) is of course different for these smooth switching profiles, but it remains strongly entangling and therefore equally suitable as a primitive two-qubit gate.

The analysis presented in the present paper has been phrased in terms of a one-dimensional array (a). However, the basic gate construction, involving two qubit bearing spins and one barrier spin, can immediately be generalised to many geometries in either two, or three dimensions. In principle one can produce a suitable structure by taking {\em any} arrangement of qubit-bearing spins, and introducing a barrier spin between each (hitherto) adjacent pair. One possible measure of the efficiency of the implementation would be the ratio of qubit-bearing spins to total number of spins, which we can denote $R_Q$. The value $R_Q=\frac{1}{2}$ corresponds to the one-dimensional arrangement (Fig 3(a)). For for a two, or higher, dimensional geometry at least some of the qubits must of course have three or more neighbors. If we restrict ourselves to considering regular structures in which every qubit has the same number of neighbors, then it is apparent that the highest possible value of $R_Q$ is $2/5$. Two arrangements which achieve this value are the hexagonal geometry Fig. 3(b), and the 3D structure illustrated in Fig. 3(c). 

In order to do better than this ratio it would be necessary for barrier spins to do `double duty' in the sense that each barrier could not be unique to a specific qubit pair. Figure (d) shows an example arrangement achieving $R_Q=3/5$ in 2D. Note (d) is the compliment of (b), i.e. the qubit and barrier roles are reversed; similarly, one could reverse Fig. 3(c) for a 3D form. In such a structure, bringing a barrier into resonance with its neigbors would initiate a three-qubit gate process - to successfully complete the gate one would require the simultaneous revival of all qubit basis states at some subsequent moment. As the number of qubits involved increases, this quickly becomes infeasible (see Appendix II), but both the three qubit gate shown in Fig. 3(d), and a four qubit variant, do appear possible \cite{SCBunpub}.
Of course, such multi-qubit gates are quite exotic and may be rather inefficient primitives for implementing algorithms.

In conclusion, we have extended the results presented in Ref. [\onlinecite{ourPRL}] in several significant respects. The first, fundamental generalization is from a pure isotropic Heisenberg interaction to a more general anisotropic interaction, including the in-plane ``XY'' interaction as a special case.
All the results presented here incorporate this generality. We have provided a proof that an interaction of this general form tends to a simple Ising interaction in the limit of far off-resonance neighbors. We have presented an analysis of the basic gate of Ref.[\onlinecite{ourPRL}] in with this general interaction, and exhibited the resulting primitive two-qubit gate. In the special case of an XY interaction, we note that the gate has an especially simple form and we provided an explicit circuit for an efficient CNOT based on this primitive. We also consider the effect of a non-abrupt switching of the Zeeman energy: by numerical simulation we demonstrate that simply varying the duration of the on-resonance phase (while the switching time remains constant) allows one to achieve the necessary revival of the barrier spins, and therefore abrupt switching is not a requirement of the scheme. Finally we have remarked upon the simplicity of generalizing to two- and three-dimensional arrays, noting that the array geometry then determines the scheme's cost in terms of the proportion of barrier spins. 

SCB wishes to acknowledge support from a Royal Society URF, and from the Foresight LINK project  ``Nanoelectronics at the Quantum Edge''.

\newpage
\noindent{\bf Appendix I: Analysis of Heisenberg Chain with Large Zeeman Discrepancies}
\bigskip 

Given the definition of $H\equiv H_{1}+H_{2}$ introduced in the main body of the paper, we can proceed to use the Trotter formula to write the time evolution operator $U(t)$ as
\begin{equation}
U(t)=\lgroup \exp{(-iH_1 t/n)}\exp{(-iH_2 t/n)}\rgroup^n\ \ \ {\rm as} \ \ \ n\rightarrow \infty. 
\label{trotter}
\end{equation}
Now we will seek to move all $H_1$ terms to the right, thus separating the Ising and non-Ising parts. Note first that since
$$
[\sigma_i^Z,\sigma_j^Z]=0\nonumber \ \ \ \ [\sigma_i^Z,\sigma_j^Z\sigma_k^Z]=0\ \ \ \ [\sigma_i^Z\sigma_j^Z,\sigma_k^Z\sigma_m^Z]=0
\nonumber
$$
we can write the following, using $\tau\equiv t/n$,
\begin{equation}
\exp(-i H_1\tau)=\lgroup\prod_j\exp(-iB_j \tau \sigma_j^Z)\rgroup\lgroup\prod_j(\exp(-i\alpha J \tau \sigma^Z_j\sigma^Z_{j+1})\rgroup 
\label{sigZfactor}
\end{equation}
and in fact we can reorder these terms as we wish.
Moreover we can use
\begin{eqnarray}
\exp(i \tau B_j \sigma_j^Z)&=&\cos(\tau B_j)1 + i\sin(\tau B_j)\sigma_j^Z\nonumber\\
\exp(i \tau J\alpha\sigma_j^Z\sigma_{j+1}^Z)&=&\cos(\tau J\alpha)1 + i\sin(\tau J \alpha)\sigma_j^Z\sigma_{j+1}^Z 
\label{sincosexpand}
\end{eqnarray}
We will also find it useful to employ
\begin{equation}
\sigma^Z\sigma^\pm=\pm\sigma^\pm,\ \ \ \sigma^\pm\sigma^Z=\mp\sigma^\pm\ \ \Rightarrow\ \ \sigma^Z\sigma^\pm=-\sigma^\pm\sigma^Z 
\label{sigmaZabsorb}
\end{equation}
where $\sigma^{\pm}$ are as defined in the main body of the paper. We will introduce a generalisation of $H_2$,
$$
H_{2}^W\equiv\frac{J}{2}(\sum_i W_j\sigma^{+}_i
\sigma^{-}_{i+1}+W_j^\dagger\sigma^{-}_i\sigma^{+}_{i+1}).
$$
where the $W_j$ are any functions involving scalar constants and $\sigma^Z_k$ for any/all $k$. Now expand
\begin{equation}
\exp(-iH^W_{2}\tau)=\sum_p^\infty \frac{1}{p!}\lgroup\frac{-iJ}{2}(\sum_j W_j\sigma^{+}_j
\sigma^{-}_{j+1}+W_j^\dagger\sigma^{-}_j\sigma^{+}_{j+1})\tau\rgroup^p
\label{powerSeries}
\end{equation}
and note the following using (\ref{sincosexpand}) and (\ref{sigmaZabsorb})
\begin{eqnarray}
&&\exp(-i\tau B_j\sigma_j^Z)\lgroup W_j\sigma^{+}_j
\sigma^{-}_{j+1}+W^\dagger_j\sigma^{-}_j\sigma^{+}_{j+1}\rgroup\nonumber\\
&&\ \ \ =\lgroup W_j\sigma^{+}_j
\sigma^{-}_{j+1}+W^\dagger_j\sigma^{-}_j\sigma^{+}_{j+1}\rgroup\exp(i \tau B_j\sigma_j^Z)\nonumber \\
&&\ \ \ =\lgroup W_j\sigma^{+}_j
\sigma^{-}_{j+1}+W^\dagger_j\sigma^{-}_j\sigma^{+}_{j+1}\rgroup\exp(2i \tau B_j\sigma_j^Z)\exp(-i \tau B_j\sigma_j^Z)\nonumber \\
&&\ \ \ =\lgroup e^{-2i\tau B_j}W_j\sigma^{+}_j
\sigma^{-}_{j+1}+e^{2i\tau B_j}W^\dagger_j\sigma^{-}_j\sigma^{+}_{j+1})\exp(-i B_j\tau\sigma_j^Z)
\nonumber
\end{eqnarray}
and similarly
\begin{eqnarray}
&&\exp(-i\tau B_{j+1}\sigma_{j+1}^Z)(W_j\sigma^{+}_j
\sigma^{-}_{j+1}+W^\dagger_j\sigma^{-}_j\sigma^{+}_{j+1})\nonumber\\
&&\ \ \ =(e^{2i\tau B_{j+1}}W_j\sigma^{+}_j
\sigma^{-}_{j+1}+e^{-2i\tau B_{j+1}}W^\dagger\sigma^{-}_j\sigma^{+}_{j+1})\exp(-i B_{j+1}\tau\sigma_j^Z).
\nonumber
\end{eqnarray}
Then
\begin{eqnarray}
&&\lgroup\prod_j(\exp(-i\tau B_j\sigma_j^Z)\rgroup(W_k\sigma^{+}_k
\sigma^{-}_{k+1}+W^\dagger_k\sigma^{-}_k\sigma^{+}_{k+1})\nonumber \\
&&\ =(\exp(i\tau\Delta_k)W_k\sigma^{+}_k
\sigma^{-}_{k+1}+\exp(-i\tau\Delta_k)W_k^\dagger\sigma^{-}_k\sigma^{+}_{k+1})\lgroup\prod_j(\exp(-i\tau B_j\sigma_j^Z)\rgroup
\nonumber
\end{eqnarray}
where $\Delta_j\equiv 2(B_{j+1}-B_{j})$. Now combining this with (\ref{sigZfactor}) and (\ref{powerSeries}) we can write
\begin{eqnarray}
&&\exp(-i H_1\tau)\exp(-i H_2^W\tau)\nonumber\\
&&\ =\lgroup\prod_j(\exp(-i\alpha J \tau\sigma^Z_j\sigma^Z_{j+1})\rgroup
\exp(-i H_2^{V}\tau)\lgroup\prod_j\exp(-iB_j \tau\sigma_j^Z)\rgroup \label{halfDone}
\end{eqnarray}
where $V_j\equiv\exp(i\tau\Delta_j)W_j$. Now we can commute the remaining left side product through to the right in a similar way. Again using (\ref{sincosexpand}) and (\ref{sigmaZabsorb}) we note that:
\begin{eqnarray}
&&\exp(-i\tau \alpha J\sigma_{j-1}^Z\sigma_j^Z)\lgroup W_j\sigma^{+}_j
\sigma^{-}_{j+1}+W^\dagger_j\sigma^{-}_j\sigma^{+}_{j+1}\rgroup\nonumber\\
&&\ \ \ =\lgroup W_j\sigma^{+}_j
\sigma^{-}_{j+1}+W^\dagger_j\sigma^{-}_j\sigma^{+}_{j+1}\rgroup\exp(i\tau \alpha J\sigma_{j-1}^Z\sigma_j^Z)\nonumber \\
&&\ \ \ =\lgroup W_j\sigma^{+}_j\sigma^{-}_{j+1}+W^\dagger_j\sigma^{-}_j\sigma^{+}_{j+1}\rgroup\exp(2i\tau \alpha J\sigma_{j-1}^Z\sigma_j^Z)\exp(-i\tau \alpha J\sigma_{j-1}^Z\sigma_j^Z)\nonumber \\
&&\ \ \ =\lgroup W_j\exp(-2i \tau \alpha J\sigma_{j-1}^Z)\sigma^{+}_j
\sigma^{-}_{j+1}+W^\dagger_j\exp(2i \tau \alpha J\sigma_{j-1}^Z)\sigma^{-}_j\sigma^{+}_{j+1}\rgroup\exp(-i\tau \alpha J\sigma_{j-1}^Z\sigma_j^Z)\nonumber
\end{eqnarray}
Similarly
\begin{eqnarray}
&&\exp(-i\tau \alpha J\sigma_{j+1}^Z\sigma_{j+2}^Z)\lgroup W_j\sigma^{+}_j
\sigma^{-}_{j+1}+W^\dagger_j\sigma^{-}_j\sigma^{+}_{j+1}\rgroup\nonumber\\
&&\ \ \ =\lgroup W_j\exp(2i \tau \alpha J\sigma_{j+2}^Z)\sigma^{+}_j
\sigma^{-}_{j+1}+W^\dagger_j\exp(-2i \tau \alpha J\sigma_{j+2}^Z)\sigma^{-}_j\sigma^{+}_{j+1}\rgroup\exp(-i\tau \alpha J\sigma_{j+1}^Z\sigma_{j+2}^Z)\nonumber
\end{eqnarray}
However, for the $\sigma_j^Z\sigma_{j+1}^Z$ term we see that
$$
[\exp(-i\tau \alpha J\sigma_{j}^Z\sigma_{j+1}^Z)\ ,\ W\sigma^{+}_j
\sigma^{-}_{j+1}+W^\dagger\sigma^{-}_j\sigma^{+}_{j+1}]=0
$$
since there is a double sign inversion. Then combining these three results we can write
\begin{eqnarray}
&&\lgroup\prod_j\exp(-i\tau \alpha J\sigma_j^Z\sigma_{j+1}^Z)\rgroup\lgroup W_k\sigma^{+}_k
\sigma^{-}_{k+1}+W^\dagger_k\sigma^{-}_k\sigma^{+}_{k+1}\rgroup\nonumber \\
&&\ =\lgroup W_k\exp(2i\tau J\alpha (\sigma^Z_{k+2}-\sigma^Z_{k-1}))\sigma^{+}_k
\sigma^{-}_{k+1}\nonumber \\
&&\ \ \ +W^\dagger_k\exp(-2i\tau J\alpha (\sigma^Z_{k+2}-\sigma^Z_{k-1}))\sigma^{-}_k\sigma^{+}_{k+1}\rgroup\lgroup\prod_j\exp(-i\tau\alpha J\sigma_j^Z\sigma_{j+1}^Z)\rgroup
\nonumber
\end{eqnarray}
Now combining this with (\ref{powerSeries})  and (\ref{halfDone}) we have
$$
\exp(-i H_1\tau)\exp(-i H_2^W\tau)=\exp(-i H_2^{Q}\tau)\exp(-i H_1\tau)
$$
where $Q_j\equiv W_j\exp(i\tau(\Delta_j+2J\alpha(\sigma^Z_{j+2}-\sigma^Z_{j-1}))$.

Now because these $Q_j$ fit within the original definition of $W_j$ (i.e. they are simply ``functions involving scalar constants and $\sigma^Z_k$ for any/all $k$''), we can just repeat the argument to commute {\em all} terms $\exp(i H_1 \tau)$ to the far left. The term originally identified as the $m^{th}$ element $\exp(i H_2 \tau)$ in the Trotter expansion (\ref{trotter}) will have $m$ terms ``$\exp(-i H_1 \tau)$'' pass `through' it, and will thus accumulate a final $Q(m)=\exp(im\frac{t}{n} (\Delta_k+2J\alpha(\sigma^Z_{k+2}-\sigma^Z_{k-1}))$. So the exact expression for the time evolution finally becomes:
\begin{eqnarray}
U(t)=R(t)\exp{(-iH_1 t)}\ \ \ \ {\rm where}& R=\{\prod_{m=1}^{m=n}\exp\lgroup\frac{-it}{n}H_R(\frac{mt}{n})\rgroup\}_{n\rightarrow\infty}\nonumber\\
H_R(\eta)=\sum_j X_j(\eta)\sigma^+_j\sigma^-_{j+1}+X^\dagger_j(\eta)\sigma^-_j\sigma^+_{j+1}\ \ &{\rm with}\ \ X_j(\eta)=\exp(i\eta(\Delta_j+2J\alpha(\sigma^Z_{j+2}-\sigma^Z_{j-1}))\nonumber
\end{eqnarray}

The right hand term in $U(t)$ is the pure Ising chain evolution we seek, but the subsequent `residual' operator $R$ is more complex. We would like to show that it tends to unity as $\delta_j\equiv J/\Delta_j\rightarrow 0$ for all $j$. Now we cannot simply integrate the terms in the product $R$ since they do not commute, and thus we cannot immediately gather the elements with a $\frac{t}{n}$ coefficient. Therefore we proceed by making the expansion $\exp A = 1+A+A^2/2+...$:

$$
\prod_{m=1}^{m=n}\exp{(-iH_R(\frac{mt}{n}) t/n)}=\prod_{m=1}^{m=n}\lgroup1-iH_R(\frac{mt}{n}) t/n+1/2(iH_R(\frac{mt}{n}) t/n))^2+...\rgroup 
$$
as ${n\rightarrow\infty}$. We cannot truncate this series since $H_R()$ is not small, but we will seek to gather and sum all terms of given order in $H_R$. We introduce 
\begin{eqnarray}
g_a&\equiv&\sum_{m=a}^n \frac{t}{n} H_R(mt/n)\nonumber\\
&\rightarrow& \int_{\eta=at/n}^t d\eta\ H_R(\eta)\nonumber\\
&=&\frac{J}{2}\sum_j\lgroup\lgroup\int_{\eta=at/n}^t d\eta  X_j(\eta)\rgroup\sigma^+_j\sigma^-_{j+1}+\lgroup\int_{\eta=at/n}^t d\eta  X_j(\eta)\rgroup\sigma^-_j\sigma^+_{j+1}\rgroup\nonumber
\end{eqnarray}

\noindent we can evaluate indefinite integrals\cite{exceptSpecial} as follows.
Defining $\Delta\equiv\Delta_0$, $\rho_j\equiv \Delta/\Delta_j$ and $\delta\equiv J/\Delta$,
$$
\frac{J}{2}\int d\eta X_j(\eta) = \frac{-i\delta}{2}x_j X_j(\eta)\ \ \ \ \ \ \ \ \frac{J}{2}\int d\eta X^\dagger_j(\eta) = \frac{i\delta}{2}x_j X^\dagger_j(\eta)
$$
where
$$
x_j\equiv\frac{\rho_j(1-8\alpha^2\delta_j^2(1+\sigma_{j-1}^Z\sigma_{J+2}^Z))(1-2\alpha\delta_j(\sigma_{j+2}^Z-\sigma_{j-1}^Z))}
{1-16\alpha^2\delta_j^2}\approx \rho_j 
$$
with the approximation holding in the limit that all $\delta_j\equiv\frac{J}{\Delta_j}\ll1$. Note that $\rho_j$, and thus $x_j$, is a modest ratio in our periodic chains (e.g. for an $ABAB..$ chain $\rho_j=(-1)^j$; for an $ABCABC..$ chain $\rho_j$ might run $1,1,-\frac{1}{2},1,1,-\frac{1}{2},..$ say). We can write the indefinite integral
\begin{eqnarray} \int d\eta\ H_R(\eta)&=&\frac{i\delta}{2}\sum_j x_j\lgroup -X_j(\eta)\sigma_j^+\sigma_{j+1}^- + X_j^\dagger(\eta)\sigma_j^-\sigma_{j+1}^+\rgroup\nonumber\\
&\equiv& \frac{i\delta}{2}K(\eta)
\label{ints}
\end{eqnarray}
Using $K()$ defined above we can write
\begin{equation}
g_a=\frac{i\delta}{2}(K(t)-K(at/n))
\label{gaeqn}
\end{equation}
Now returning to the expansion, the lowest order in $H_R$ is of course $1$, and the sum of all terms of $1^{st}$ order in $H_R$ is precisely $-ig_0=\frac{1}{2}\delta\{K(t)-K(0))\}$. Thus so far we are seeing the anticipated behavior: the `residual' part of the dynamics, after the Ising-like behavior is allowed for, appears to vanish with $\delta$. However, since we are using an expansion in $H_R()$, where $H_R()$ is not small, we should evaluate and sum the higher terms. Let us use the symbol $S_N$ to represent the sum of terms of order $H_R()^N$; then we have already found $S_1=\frac{1}{2}\delta\{K(t)-K(0)\}$, and 
\begin{eqnarray}
S_2&=&(-i\frac{t}{n})^2\sum_{m=1}^n \{\frac{1}{2}(H_R(mt/n))^2+H_R(mt/n)\sum_{p>m} H_R(pt/2)\} \nonumber \\
&=&-(\frac{t}{n})^2\sum_{m=1}^n \{-\frac{1}{2}(H_R(mt/n))^2+H_R(mt/n)\sum_{p\geq m} H_R(pt/2)\} \nonumber 
\end{eqnarray}
now the factor $(t/n)^2$ causes the first term here to vanish in the limit $n\rightarrow\infty$, since it contains only $n$ terms each of Order($H_R()\sim J$). For the second term
\begin{eqnarray}
S_2=-\int_0^t H_R(\eta)\lgroup \int_\eta^t H_R(\eta) d \eta\rgroup d\eta
\end{eqnarray}
but the inner integral is given by (\ref{gaeqn}) so that
\begin{equation}
S_2=\frac{-i\delta}{2}\int_0^t H_R(\eta)\lgroup K(t)-K(\eta) \rgroup d\eta
\label{order2terms}
\end{equation}
This integral can be fully evaluated\cite{SCBunpub} but the key point is that it can already be seen to be of order $\delta$ (or less). Note that the $\delta\equiv\frac{J}{\Delta}$ factor {\em cannot} be absorbed by the remaining integral since the variable $\Delta$ occurs only as a phase $\sim\exp(i\Delta\tau)$. Thus in expanding and evaluating the integral we will see some terms with an additional factor of $\frac{1}{\Delta}$, and in the special case that a term exhibits cancellation of the $\Delta$ elements in the phase we would apply a factor of order unity - but we can never introduce a factor of $\Delta$. 

Generalizing this observation we can consider $S_N$. This involves terms of the form $H_R(m_1t/n)H_R(m_2t/n)...H_R(m_Nt/n)$ for some set of integers $m_1\geq m_2\geq...m_N$. By the same reasoning above, we can neglect terms where two or more of the $m_i$ are the same value, since they collectively constitute a negligible portion $\frac{1}{n}$ of the sum as $n\rightarrow\infty$. Then we find 
\begin{eqnarray}
S_N&=&(-i)^N\int_0^t H_R(\xi_1) \int_{\xi 1}^t H_R(\xi_2)\int_{\xi 2}^t ....\int_{\xi N-1}^t H_R(\xi_N)\ d\xi_1\ d\xi_2\ ... d\xi_N\nonumber\\
&=&\frac{(i)^{N-1}}{2}\delta\int_0^t H_R(\xi_1) \int_{\xi 1}^t H_R(\xi_2) \int_{\xi 2}^t ....\int_{\xi_{N-2}}^t (K(t)-K(\xi_{N-1}))\ d\xi_1\ d\xi_2\ ... d\xi_{N-1}\nonumber
\end{eqnarray}
And as before we can argue that although the remaining $N-1$ integrals may produce additional factors of $1/\Delta$, they cannot absorb any. Thus the factor $\delta$ will remain and we can conclude that {\bf all terms in the expansion $S_n$ ($n\geq 1$) are of order $\delta$ or less.} Therefore the time evolution operator is 
$$
U(t)=(1-\delta P(t))\exp{(-iH_1 t)}
$$
for some finite operator $P(t)$ whose magnitude does not increase with $\Delta$. This is the result presented in the main body of the paper.
\bigskip

\noindent{\bf Appendix II: Regarding Revivals}

In the discussion of two and three dimensional arrays, we stated that it will be difficult to achieve the crucial simultaneous `revivals' for multi-qubit gates involving more than a few qubits. Of course, one can observe that if we choose {\em any} detuning $A-B$ for which the revival periods of the various qubit basis states are related by irrational factors (i.e. the general case), then there will eventually be a complete revival to any desired accuracy (although never perfect). However one would typically need to wait an extremely long time for the level of precision required for QC and therefore this type of revival is not a practical choice. Instead we seek to arrange rapid revivals by looking for values of the detuning (and potentially, other parameters) such that the various revival periods are all related by small rational factors. Fulfilling this condition will become unfeasible as the number of qubits increases.


\begin{references}
\bibitem{sin4function} The function here is a symmetric two part composite. Defining $\tau\equiv t-t_0$, the Zeeman shift is introduced (`switched on') in the period $0<\tau<t_\Delta$ by a function of the form $-\sin^4(\phi\tau)$ for $\tau<\frac{t_\Delta}{2}$, and $\sin^4(\phi(\tau-t_\Delta)-1$ for $\tau>\frac{t_\Delta}{2}$, where $\phi=2\arctan(2^{-1/4})/(t_\Delta)$.
\bibitem{completeR} For each profile we performed a series of numerical simulations, manually adjusting duration until we obtained revivals that were perfect to within an error probability of about $1$ part in $10^6$; apparently one could continue to refine the value arbitrarily.
\bibitem{exceptSpecial} The integral is valid provided $4J\alpha\neq \Delta_j$ $\forall \Delta_j$, which is of course the case since we are interested in the $\alpha J\ll \Delta_j$ limit.
\end{references}
\end{document}